\documentclass[12pt,letterpaper,oneside]{article}
\usepackage{amssymb,tabularx,multirow,amsmath,natbib, epsfig, threeparttable, amstext, subfigure,xcolor}
\usepackage{pdflscape}
\usepackage{afterpage}
\usepackage{capt-of}% or use the larger `caption` package
\usepackage{rotating}
\usepackage[makeroom]{cancel}
\usepackage{bm}

\newcommand{\blind}{0}

\newcommand{\rf}{\vskip .1in\par\sloppy\hangindent=1pc\hangafter=1
                 \noindent}

\def\spacingset#1{\renewcommand{\baselinestretch}%
{#1}\small\normalsize} \spacingset{1}

\usepackage[dvips,top=1.2in,bottom=0.65in,left=1.15in,right=1.15in,includefoot]{geometry}

\usepackage[margin=20pt,labelfont={bf,small},textfont={it,small},indention=0.5cm]{caption}

\begin{document}
\if0\blind
{
  \title{\bf Animal Movement Models with Mechanistic Selection Functions}
  \author{Mevin B. Hooten\thanks{
    corresponding author; hooten@rams.colostate.edu}\hspace{.2cm}\\
    U.S. Geological Survey \\ Colorado Cooperative Fish and Wildlife Research Unit \\ Department of Fish, Wildlife, and Conservation Biology \\ Department of Statistics \\ Colorado State University \\
    and \\
    Xinyi Lu \\
    Department of Statistics \\ Colorado State University \\
    and \\
    Martha J. Garlick \\
    Department of Mathematics and Computer Science \\ South Dakota School of Mines and Technology \\
    and \\
    James A. Powell \\
    Department of Mathematics and Statistics \\ Utah State University} 
  \maketitle
} \fi

\if1\blind
{
  \bigskip
  \bigskip
  \bigskip
  \begin{center}
    {\LARGE\bf Inference on Dynamics in Animal Movement Models}
\end{center}
  \medskip
} \fi

\bigskip
\begin{abstract}
A suite of statistical methods are used to study animal movement.  Most of these methods treat animal telemetry data in one of three ways:  as discrete processes, as continuous processes, or as point processes.  We briefly review each of these approaches and then focus in on the latter.  In the context of point processes, so-called resource selection analyses are among the most common way to statistically treat animal telemetry data.  However, most resource selection analyses provide inference based on approximations of point process models.  The forms of these models have been limited to a few types of specifications that provide inference about relative resource use and, less commonly, probability of use.  For more general spatio-temporal point process models, the most common type of analysis often proceeds with a data augmentation approach that is used to create a binary data set that can be analyzed with conditional logistic regression.  We show that the conditional logistic regression likelihood can be generalized to accommodate a variety of alternative specifications related to resource selection.  We then provide an example of a case where a spatio-temporal point process model coincides with that implied by a mechanistic model for movement expressed as a partial differential equation derived from first principles of movement.  We demonstrate that inference from this form of point process model is intuitive (and could be useful for management and conservation) by analyzing a set of telemetry data from a mountain lion in Colorado, USA, to understand the effects of spatially explicit environmental conditions on movement behavior of this species.     
\end{abstract}

\noindent%
{\it Keywords:}  partial differential equation, point process, resource selection function, step selection function 
%\vfill

%\newpage
\spacingset{1.45} % DON'T change the spacing!

\section{Introduction}
The dynamics associated with animals moving in complex environments are critical to the natural function of the individual, species, and ecosystem in which animals reside (Nathan et al., 2008).  To study movement dynamics, a wide variety of statistical models have been developed to analyze animal telemetry data (Hooten et al., 2017).  In general, these statistical models are based on perspectives of movement and data generating mechanisms that fall into 3 main classes:  discrete-time processes, continuous-time processes, and point processes (Hooten and Johnson, 2019).  These same three types of processes are also the main subject of study in spatial statistics (Cressie, 1993) and are thus familiar to most spatial statisticians.  Of course, when considered in time explicitly, the trajectories of moving animals are spatio-temporal processes and it is natural to account for temporal dependence in animal movement models just as we would when modeling other dynamic processes (Wikle and Cressie, 2011).  The temporal dependence in telemetry data can provide insights about important biological and ecological dynamic processes (Hooten et al., 2019).   

In what follows, we review the three main classes of statistical models that are used to analyze animal telemetry data.  We then delve more deeply into point process models and the commonly used implementations of them.  Finally, we show that an unusual form of point process model arises naturally as a result of a partial differential equation that describes the movement of animals based on mechanistic first principles.  We demonstrate how to fit the resulting point process model to data using a popular conditional regression procedure.       

\subsection{Overview of Statistical Models for Animal Movement}

We assume the true position of an individual animal is measured and expressed as $\mathbf{s}(t_i)$ for time $t_i$ and with support $\mathbf{s}(t_i)\in {\cal S}$ (often a subset of two-dimensional geographic space) for $i=1,\ldots,n$ observation times.  A variety of devices and approaches are used to observe the animal position $\mathbf{s}(t_i)$ (e.g., Cooke et al., 2004) and the associated measurement error can be accounted for in a hierarchical framework (e.g., Brost et al., 2015).  For the purposes of this exposition, we assume the measurement error is small enough to be negligible, such as that arising from high-quality global positioning system (GPS) telemetry devices (e.g., Cagnacci et al., 2010).

Discrete-time models for movement most closely follow methods used in time series analysis.  For example, following Hooten et al. (2017), a temporal vector autoregressive model for the position $\mathbf{s}(t_i)$ can be expressed as 
\begin{equation} 
  \mathbf{s}(t_i) = (\mathbf{I}-\mathbf{A})\mathbf{c}+\mathbf{A}\mathbf{s}(t_{i-1})+\boldsymbol\varepsilon(t_i) \;, 
  \label{eq:var}
\end{equation} 
\noindent where, $\mathbf{I}$ is a $2\times 2$ identity matrix and, for now, we assume that the time lag between observations $\Delta t= t_i - t_{i-1}$ is constant and the error term $\boldsymbol\varepsilon(t_i) \sim \text{N}(\mathbf{0},\boldsymbol\Sigma)$.  In the example model in (\ref{eq:var}), the $2\times 1$ vector $\mathbf{c}$ represents the activity center for the animal in geographic space and the $2\times 2$ propagator matrix $\mathbf{A}$ controls the dynamics of the discrete-time trajectory.  Anderson-Sprecher and Ledolter (1991) described this form of discrete-time movement model and various extensions.    

Similar models were developed for continuous-time movement and they are often based on Weiner processes that are represented as stochastic differential equations (SDEs).  Both the discrete- and continuous-time models have ``integrated'' forms that account for additional smoothness in the trajectories (e.g., Jonsen et al., 2005; Johnson et al., 2008b; Hooten and Johnson, 2017).  

Accounting for heterogeneity in movement dynamics has been a strong theme in contemporary movement modeling (Hooten et al., 2018).  Morales et al. (2004) proposed representing the velocity associated with the changes in position in polar coordinates and using the resulting step-length and turning angle as response variables in a variety of statistical mixture models.  These mixture models cluster portions of the animal trajectories into interpretable groups that can be associated with distinct animal behaviors (e.g., resting, foraging, transit).  This class of discrete-time movement models is now most commonly implemented using a hidden Markov model framework (Zucchini et al., 2009) and a variety of software exists to fit these models to data (e.g., McClintock and Michelot, 2018).   

The final class of animal movement models is based on point processes from spatial statistics.  However, it appears that much of the methodology was developed somewhat independently in wildlife ecology (Manly et al., 2002) because, while the model forms are the same, the terminology varies across fields.  Point process models treat a set of points as the response variables and account for heterogeneity in the space from which the points arose using a spatially varying density function.  The density function for a point process is often expressed as a weighted distribution (Patil and Rao, 1977) where, for example, a trajectory observation arises as 
\begin{equation}   
  \mathbf{s}(t_i) \sim \frac{g(\mathbf{s}(t_i),\boldsymbol\theta)}{\int_{\cal S} g(\mathbf{s},\boldsymbol\theta)d\mathbf{s}} \;, 
  \label{eq:spp}
\end{equation}   
\noindent where the function $g$ is non-negative and referred to as the resource selection function (RSF).  Thus, point process models for telemetry data are called RSF models by wildlife ecologists.   

The denominator in (\ref{eq:spp}) presents a challenge for implementing point process models because the integral is often intractable.  Thus, logistic (or Poisson) regression procedures have been developed to approximate the likelihood and fit these models to data using existing statistical software.  We elaborate on these approximation methods for fitting point process models in the next section.     

Point process models are popular among wildlife ecologists because they are easy to implement and provide inference about the selection of ``resources'' (represented as covariates in the RSF) thereby connecting the moving animal to its environment.  Continuous-time models based on SDEs have also been developed to provide inference about how individual movement corresponds to changes in the environment (e.g., Preisler et al., 2004; Brillinger, 2010; Hooten et al., 2019), but they can be challenging to implement by practitioners and are thus less popular.      

When considering the temporal dynamics of a movement trajectory, spatio-temporal point process models (STPPs) have been adapted to the animal movement setting (e.g., Johnson et al., 2008a; Johnson et al., 2013; Brost et al., 2015).  In some cases, STPPs have been implemented using conditional logistic regression procedures (Fortin et al., 2005); in the animal movement setting, the use of conditional logistic regression to fit approximate point process models is often referred to as step-selection analysis.  We focus on spatio-temporal point process models and their implementation in what follows.      

\section{Methods}
\subsection{STPPs for Animal Movement}
Conditioning on the total number of observations ($n$), a STPP model for the observed positions $\mathbf{s}(t_i)$ is often expressed using a weighted distribution (Patil and Rao, 1977) representation as  
\begin{equation}
  [\mathbf{s}(t_i) | \mathbf{s}(t_{i-1}), \boldsymbol\theta] = \frac{g(\mathbf{w}(\mathbf{s}(t_i)),\boldsymbol\theta)f_i(\mathbf{s}(t_i)|\mathbf{s}(t_{i-1}))}{\int_{\cal S}g(\mathbf{w}(\mathbf{s}),\boldsymbol\theta)f_i(\mathbf{s}|\mathbf{s}(t_{i-1})) d\mathbf{s}} \;,
  \label{eq:stpp}
\end{equation}
\noindent where the bracket notation `$[\cdot]$' corresponds to a probability distribution (Gelfand and Smith, 1990) and $\mathbf{w}(\mathbf{s})$ represents a vector of covariates at position $\mathbf{s}$.  The functions $g$ and $f$ in (\ref{eq:stpp}) are non-negative and often referred to as the ``selection'' (as in the RSF models described previously) and ``availability'' functions, respectively, in the animal ecology literature (Hooten et al., 2017).  When the availability function $f_i(\mathbf{s}(t_i)|\mathbf{s}(t_{i-1}))$ is specified as a uniform probability density function over ${\cal S}$, then the model in (\ref{eq:stpp}) simplifies to the RSF model previously described (Manly et al., 2002; Johnson et al., 2006).  The traditional RSF model assumes that the observed positions arise conditionally independent of one another.  Most commonly, the RSF model is fit using a data augmentation strategy where a set of indicators serve as the response variable in a binary regression.  When the RSF is specified as $g(\mathbf{w}(\mathbf{s}(t_i)),\boldsymbol\theta)\equiv e^{\mathbf{w}'(\mathbf{s}(t_i))\boldsymbol\theta}$, logistic regression is implied and can be implemented using a variety of software.    

In the case where the availability function in (\ref{eq:stpp}) is time dependent, such as when the telemetry data are collected at a high temporal resolution, the likelihood associated with (\ref{eq:stpp}) can be approximated using a similar data augmentation strategy and conditional logistic regression (Breslow and Day, 1980; Boyce, 2006).  In this approach, the analyst creates a binary data set comprised of a single one ($y_{i1}=1$) representing a ``presence'' associated with the observed position $\mathbf{s}(t_i)$, and zeros ($y_{ij}=0$ for $j=2,\ldots,J$) for a set of locations simulated from the availability function at each time step $t_i$ based on the position at the previous time $t_{i-1}$ (i.e., $J-1$ positions $\mathbf{s}_{ij}$ where the individual did not move are simulated from the normalized availability function $f_i(\mathbf{s}|\mathbf{s}(t_{i-1}))$ at each time step).  Using the associated covariate values at the observed and simulated availability positions, the likelihood for the STPP model in (\ref{eq:stpp}) can be approximated with that resulting from the binary data model $y_{ij}\sim \text{Bern}(p_{ij})$ when $J\rightarrow\infty$, where $\text{logit}(p_{ij})=\beta_{0,i}+\log(g(\mathbf{w}'_{ij}\boldsymbol\theta))$.  The step-specific intercepts $\beta_{0,i}$ for $i=1,\ldots,n$ account for the changing availability at each time step when making inference on $\boldsymbol\theta$.  In this approach, the function $g$ in (\ref{eq:stpp}) is referred to as a step-selection function (SSF) and the associated analysis is a step-selection analysis (Fortin et al., 2005).  In practice, the constraint $\sum_{j=1}^J y_{ij} = 1$ allows us to derive a conditional likelihood that we can maximize using standard statistical software (often the same software that is used to fit Cox proportional hazard models to survival data). 

Several important notes are relevant to this practice.  First, it is possible, but not common, to maximize the original point process likelihood associated with (\ref{eq:stpp}) directly (e.g., Johnson et al., 2008a; Johnson et al., 2013; Brost et al., 2015).  However, because the integral in the denominator of (\ref{eq:stpp}) can be costly to compute in an iterative algorithm and because the availability function $f_i(\mathbf{s}(t_i)|\mathbf{s}(t_{i-1}))$ may be complicated, most practitioners use the data to estimate the availability distribution \emph{a priori} and sample from a normalized version of it, and then use conditional logistic regression with available software (e.g., Signer et al., 2019).  When Bayesian methods are used, this results in an empirical Bayes procedure that often provides a good representation of the true model.  Such implementations are often justified by practitioner claims that the RSF (or SSF) is the main focus of their inference and the availability function exists only to account for additional temporal dependence.          

In what follows, we present a derivation of the conditional likelihood associated with the empirical Bayes approach to fitting an STPP with a general selection function $g(\mathbf{w}(\mathbf{s}(t_i)),\boldsymbol\theta)$.  We then use the resulting likelihood in a Bayesian model that has connections to the same mechanisms that have been used to describe spatio-temporal population dynamics.   

\subsection{Conditional Regression Procedure}
The approaches to resource-selection and step-selection analyses described above typically rely on a specification of the selection function as $g(\mathbf{w}(\mathbf{s}(t_i)),\boldsymbol\theta)\equiv e^{\mathbf{w}'(\mathbf{s}(t_i))\boldsymbol\theta}$.  However, when an intercept ($\theta_0$) is included in the selection function such that $\mathbf{w}'(\mathbf{s}(t_i))\boldsymbol\theta=\theta_0+\theta_1 \mathbf{x}_{1}+\ldots+\theta_{p-1} \mathbf{x}_{p-1}$, it cancels in the RSF and SSF likelihoods and limits the inference to relative selection only (Manly et al., 2002).  In such cases, the researcher can only say that the individual selects for a resource more (or less) than another resource; they cannot infer the absolute probability of selection (Lele and Keim, 2006).  This fact has led some to argue for the use of resource selection probability functions (RSPFs) specified in such a way that $g(\mathbf{w}(\mathbf{s}(t_i)),\boldsymbol\theta)$ is a probability function such as the inverse logit or probit that are bounded below by zero and above by one (Lele and Keim, 2006; Lele, 2009; Solymos and Lele, 2016).               

In the sections that follow, we highlight other forms of selection functions, for $g(\mathbf{w}(\mathbf{s}(t_i)),\boldsymbol\theta)>0$, that ecologists may wish to consider for inference.  Thus, in the case of the general SSF model in (\ref{eq:stpp}), a similar logistic regression procedure to that described in the previous section can be considered where 
\begin{align} 
  y_{ij} &\sim \text{Bern}(\phi_{ij}) \;, \\ 
  \text{logit}(\phi_{ij}) &= \beta_{0,i}+\log(g(\mathbf{w}_{ij},\boldsymbol\theta)) \;,
\end{align} 
\noindent for $i=2,\ldots,n$ steps, where $y_{i1}=1$ and $y_{ij}=0$ for $j=2,\ldots,J$ availability samples, and where $\mathbf{w}_{ij}$ represents the set of covariate values at the $j$th availability position for step $i$.  Under this logistic regression procedure, the likelihood component associated with step $i$ is 
\begin{align} 
  [\mathbf{y}_i|\beta_{0,i},\boldsymbol\theta] &=\prod_{j=1}^J \phi_{ij}^{y_{ij}}(1-\phi_{ij})^{1-y_{ij}} \;, \\
  &= \prod_{j=1}^J \left(\frac{e^{\beta_{0,i}+\log(g(\mathbf{w}_{ij},\boldsymbol\theta))}}{1+e^{\beta_{0,i}+\log(g(\mathbf{w}_{ij},\boldsymbol\theta))}}\right)^{y_{ij}}\left(\frac{1}{1+e^{\beta_{0,i}+\log(g(\mathbf{w}_{ij},\boldsymbol\theta))}}\right)^{1-y_{ij}} \;, \\
  &= \frac{e^{\beta_{0,i}\sum_{j=1}^J y_{ij}+\sum_{j=1}^J \log(g(\mathbf{w}_{ij}),\boldsymbol\theta)) y_{ij}}}{\left(1+e^{\beta_{0,i}+\log(g(\mathbf{w}_{ij},\boldsymbol\theta))}\right)^J} \;.
\end{align} 
When we account for the known constraint $\sum_{j=1}^J y_{ij}=1$ for all $i=2,\ldots,n$, the resulting conditional likelihood component becomes   
\begin{align}
  \left[\mathbf{y}_i\bigg\vert\boldsymbol\theta,\sum_{j=1}^J y_{ij}=1\right] &= \frac{[\mathbf{y}_i|\beta_{0,i},\boldsymbol\theta]}{\sum_{\tilde{\mathbf{y}}'_i\mathbf{1}=1}[\tilde{\mathbf{y}}_i|\beta_{0,i},\boldsymbol\theta]} \;, \\
  &= \frac{e^{\sum_{j=1}^J \log(g(\mathbf{w}_{ij},\boldsymbol\theta)) y_{ij}}}{\sum_{\tilde{\mathbf{y}}'_i\mathbf{1}=1} e^{\sum_{j=1}^J \log(g(\mathbf{w}_{ij},\boldsymbol\theta)) \tilde{y}_{ij}}} \;,
\end{align}
\noindent because the term $e^{\beta_{0,i}\sum_{j=1}^J y_{ij}}$ and the denominator of $[\mathbf{y}_i|\beta_{0,i},\boldsymbol\theta]$ cancel.  Note that the sum $\tilde{\mathbf{y}}'_i\mathbf{1}=1$ includes all possible arrangements of binary data for step $i$.  Thus, the complete conditional likelihood for all steps $i=2,\ldots,n$ is 
\begin{equation}
  \left[\mathbf{Y}\bigg\vert\boldsymbol\theta,\left\{\sum_{j=1}^J y_{ij} = 1, \forall i\right\} \right] = \prod_{i=2}^n \frac{g(\mathbf{w}_{i1},\boldsymbol\theta)}{\sum_{j=1}^J g(\mathbf{w}_{ij},\boldsymbol\theta)}
  \label{eq:condlik}
\end{equation}
\noindent because only $y_{i1}=1$ for step $i$ (the rest of $y_{ij}=0$ for $j=2,\ldots,J$).

The binary regression model with intercept terms that vary with step is not by itself a generative model for the data, but is nonetheless used as a means to achieve a likelihood that approximates the STPP model.  An alternative approach would be to use a multinomial model where the binary data for each step are restricted by the multinomial distribution to contain only a single value of one (i.e., $\mathbf{y}_i \sim \text{MN}(1,\boldsymbol\phi_i)$).  We show that the multinomial approach results in the same likelihood as the conditional logistic regression procedure in Appendix A. 

Thus, regardless of whether a Bernoulli or multinomial model for the augmented data is assumed, a form of conditional logistic regression can be used to fit the STPP model using a SSF of choice as long as we assume or empirically estimate the availability function and obtain a large set (i.e., $J\rightarrow\infty$; Northrup et al., 2013) of positions from it at each time step to construct the augmented binary data set $\mathbf{Y}$.       

\subsection{Partial Differential Equation for Movement}
Turchin (1998) showed that a form of partial differential equation (PDE) called the Fokker-Planck equation can be derived from a discrete-time Lagrangian movement model.  The procedure for deriving the movement-based Fokker-Planck equation involves expanding a set of movement and residence probabilities in a Taylor series, truncating higher-order terms, and rearranging to yield a PDE with motility parameters appearing inside the second spatial derivative (Hooten and Wikle, 2010; Hooten et al., 2013).  While advection-diffusion PDEs have been used in environmental science for decades (Wikle and Hooten, 2010; Cressie and Wikle, 2011), their use in statistical models for population-level animal movement has also become popular recently (e.g., Wikle, 2003; Hooten and Wikle, 2008).  The two-dimensional diffusion form of the Fokker-Planck PDE is
\begin{equation}
  \frac{\partial p(\mathbf{s},t)}{\partial t}=\left(\frac{\partial^2}{\partial s^2_1}+\frac{\partial^2}{\partial s^2_2}\right) \delta(\mathbf{s}) p(\mathbf{s},t) \;,  
  \label{eq:EDE}  
\end{equation}
\noindent for probability of presence $p(\mathbf{s},t)$, and is also called the ecological diffusion equation (EDE).  The EDE in (\ref{eq:EDE}) involves the motility function $\delta(\mathbf{s})=\frac{\Delta s^2}{4\Delta t}\psi(\mathbf{s})$ which relates to the movement probability $\psi(\mathbf{s})$ from the original Lagrangian model with spatial grain $\Delta s^2$ and temporal resolution $\Delta t$.  The EDE is especially relevant for modeling movement because it can be derived from first principles of individual-level movement and results in residence times (i.e., the length of time an individual resides in an area before moving) that are realistically related to landscape pattern (Powell and Zimmerman, 2004; Garlick et al., 2011; Garlick et al., 2014).       

Multiplying the presence probability in the EDE (\ref{eq:EDE}) by population abundance yields a spatio-temporal model for population intensity (Hooten et al., 2013).  The resulting population-level models have been used in a variety of statistical implementations and ecological applications (Hefley et al., 2017; Williams et al., 2017; Lu et al., In Press).  Only recently has the EDE been considered in an individual animal movement context.  Garlick et al. (In Review) showed that a fundamental solution of the ``homogenized'' EDE (Appendix B) has the form 
\begin{equation}  
  [\mathbf{s}(t_i)|\mathbf{s}(t_{i-1})] \propto \frac{1}{\delta(\mathbf{s}(t_{i}))\Delta t_i}e^{-\frac{1}{2}(\mathbf{s}(t_i)-\mathbf{s}(t_{i-1}))'(2\bar{\delta}(t_i)\Delta t_i\mathbf{I})^{-1}(\mathbf{s}(t_i)-\mathbf{s}(t_{i-1}))} \;,
  \label{eq:pss}
\end{equation}  
\noindent where the term $\bar{\delta}(t_i)$ in (\ref{eq:pss}) is a local harmonic mean of motility $\delta(t)$ that arises naturally from the homogenization method.  Homogenization is a multiscale approximation technique that can be used with certain classes of PDEs to make them more computationally efficient to solve numerically (Hooten et al., 2013).  In the case of the EDE, homogenization also facilitates the fundamental solution in (\ref{eq:pss}) which can be used as a statistical model for animal trajectories.   

The critical aspect of the fundamental solution to the EDE in (\ref{eq:pss}) that makes it relevant to our review of STPPs is that it takes the form of the point process model in (\ref{eq:stpp}).  If we define the selection function from (\ref{eq:stpp}) as 
\begin{equation}
  g(\mathbf{w}(\mathbf{s}(t_i)),\boldsymbol\theta)=\frac{1}{\delta(\mathbf{s}(t_{i}))\Delta t_i} \;,
  \label{eq:selfcn}
\end{equation}
\noindent and the availability function as 
\begin{equation}
  f(\mathbf{s}(t_i)|\mathbf{s}(t_{i-1}))\propto e^{-\frac{1}{2}(\mathbf{s}(t_i)-\mathbf{s}(t_{i-1}))'(2\bar{\delta}(t_i)\Delta t_i\mathbf{I})^{-1}(\mathbf{s}(t_i)-\mathbf{s}(t_{i-1}))} \;,
  \label{eq:avlfcn}
\end{equation}
\noindent then the fundamental solution (\ref{eq:pss}) to the homogenized EDE is a member of the class of statistical point process models based on the weighted distribution specification in (\ref{eq:stpp}).  It is easily shown that the availability function (\ref{eq:avlfcn}) is an unnormalized multivariate normal density function for $\mathbf{s}(t_i)$ that lends itself to straightforward stochastic simulation.  However, the selection function (\ref{eq:selfcn}) is notably different than those used in former developments of RSFs and SSFs.  Using the relationship between the motility function $\delta(\mathbf{s})$ and the movement probability $\psi(\mathbf{s})$, we can reduce the dimensionality for statistical estimation by linking the movement probability to a set of environmental covariates $\mathbf{w}(\mathbf{s})$ via $\text{logit}(\psi(\mathbf{s}))=\mathbf{w}'(\mathbf{s})\boldsymbol\theta$.     

Following Turchin (1998), Garlick et al. (In Review) showed that the homogenized motility coefficients $\bar{\delta}(t_i)$ could be pre-estimated with a temporal moving average of the original telemetry data 
\begin{equation}
  \bar{\delta}(t_i)\approx \sum_{t_j\sim t_i} \frac{(\mathbf{s}(t_j)-\mathbf{s}(t_{j-1}))'(\mathbf{s}(t_j)-\mathbf{s}(t_{j-1}))}{4 n_i\Delta t_j} \;,
  \label{eq:deltabar} 
\end{equation}
\noindent where, $t_j\sim t_i$ indicates the set of times $t_j$ that are considered temporally close to $t_i$ and $n_i$ is the size of the set $t_j\sim t_i$.  When used in the STPP likelihood implied by the EDE, the inference on selection parameters $\boldsymbol\theta$ was robust to the pre-estimation of $\bar{\delta}(t_i)$.   

\subsection{Implementation for EDE Point Process Model}
One benefit of the EDE-based STPP model is that, like the RSPF approaches (e.g., Lele, 2006), an intercept can be included in the linear ($\mathbf{w}'(\mathbf{s})\boldsymbol\theta$) component of the selection function (\ref{eq:selfcn}) and we can obtain inference on the relationship between environmental covariates and the movement probability (and hence motility) implied by the EDE.  Additionally, because the movement probabilities are inversely related to residence time by 
\begin{equation}
  r(\mathbf{s})=\frac{4\Delta t}{\text{logit}^{-1}(\mathbf{w}'(\mathbf{s})\boldsymbol\theta)} \;, 
  \label{eq:rs}
\end{equation}
\noindent we can obtain spatially-explicit maps of estimated time spent $r(\mathbf{s})$ in a spatial region with area $\Delta s^2$ that can be used by practitioners to improve the understanding, conservation, and management of wildlife.  A knowledge of residence time is particularly important for threatened and endangered species with high site fidelity and/or philopatry (e.g., Gerber et al., 2019) and for cases where the environment may become pathogenic with increased exposure (e.g., Garlick et al., 2014).  

To connect the implementation of the EDE-based point process model to the procedures most commonly used to estimate RSFs and SSFs in the wildlife ecology literature, we used a conditional regression approach as outlined in Section 2.2.  We also developed algorithms to fit the model using Bayesian methods for two reasons:  1) to allow for straightforward inference on nonlinear derived quantities of model parameters such as $r(\mathbf{s})$ in (\ref{eq:rs}) and 2) because, like Lele (2009) in his development of RSPFs, we also found evidence of non-Gaussian and asymmetric shaped multivariate likelihood surfaces and posterior distributions for $\boldsymbol\theta$.  

Thus, to fit the EDE-based point process models, we first created an augmented binary data set consisting of a single value of one for $y_{i1}=1$ and $J-1$ zeros ($y_{ij}$ for $j=2,\ldots,J$) for which the positions were drawn randomly from the multivariate availability distribution in (\ref{eq:avlfcn}).  For each position, the associated covariate values $\mathbf{w}_{ij}$ were extracted from the environmental data sets as is standard practice in step-selection analyses.  We then used the conditional likelihood we derived in (\ref{eq:condlik}) as the approximate STPP model (that becomes exact as $J\rightarrow \infty$) and multivariate normal prior for regression coefficients $\boldsymbol\theta\sim \text{N}(\boldsymbol\mu_\theta,\boldsymbol\Sigma_\theta)$.  

We fit the resulting Bayesian EDE-based point process model using a custom Hamiltonian Monte Carlo (HMC) algorithm to accommodate situations with correlated joint posterior distributions.  Our HMC algorithm (details in Appendix C) performed well in our simulations and real data analyses.   

\section{Application}
\begin{figure}[htp]
  \centering
  \includegraphics[width=4in, angle=0]{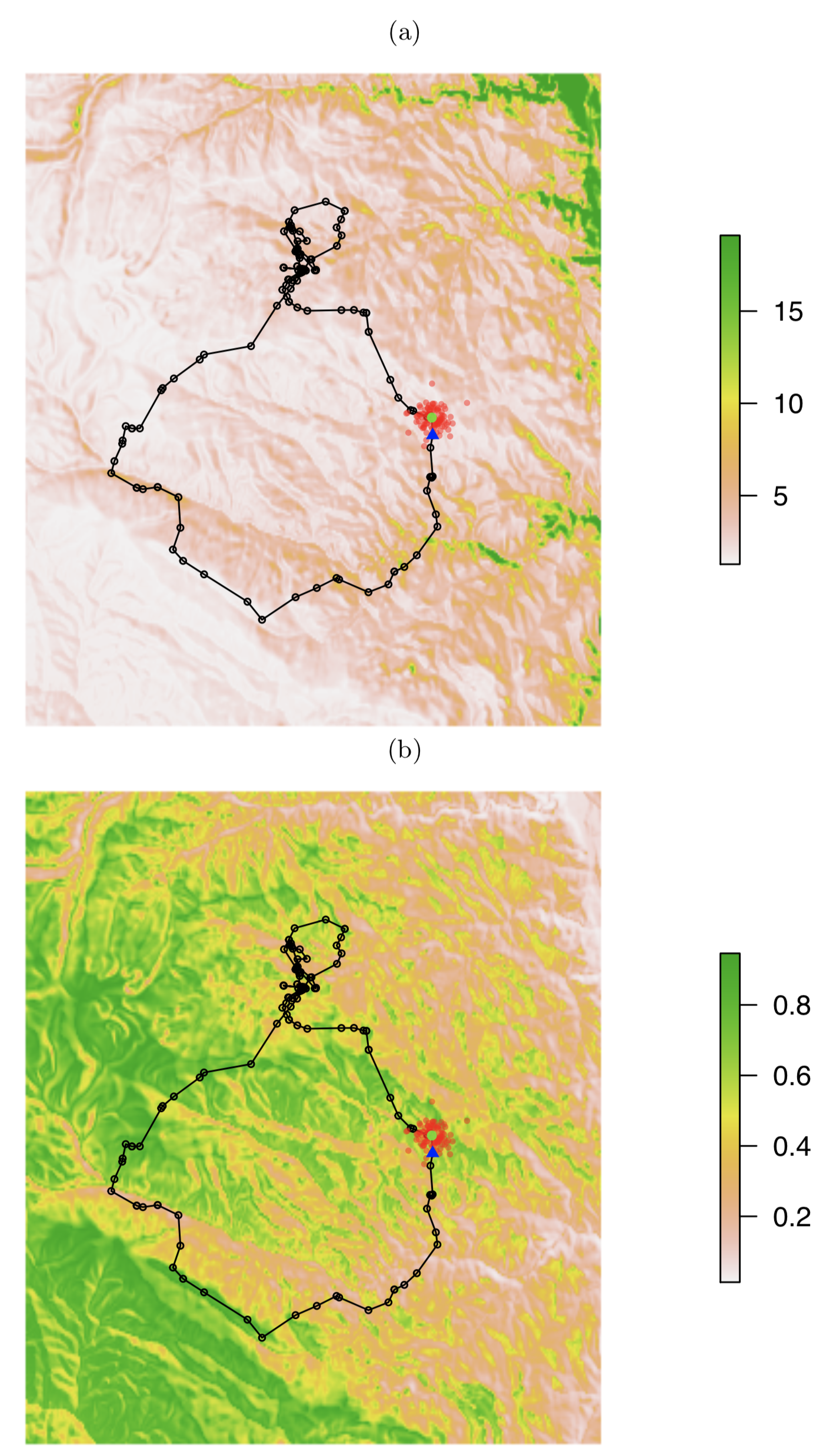}
  \caption{A total of 150 observed GPS positions (black points), connected by lines, for a single mountain lion spanning approximately 2.5 weeks in Colorado, USA.  For illustration, red points represent the availability sample ($J=100$) for step 49 based on the 49th observed GPS position shown as green point. The blue triangle represents the 50th observed GPS position. Background images show a) posterior mean residence time $r(\mathbf{s})$ in hours per hectare and b) movement probability $\psi(\mathbf{s})$ per hectare.}  
  \label{fig:ml_rs}
\end{figure}

\begin{figure}[htp]
  \centering
  \includegraphics[width=6.5in, angle=0]{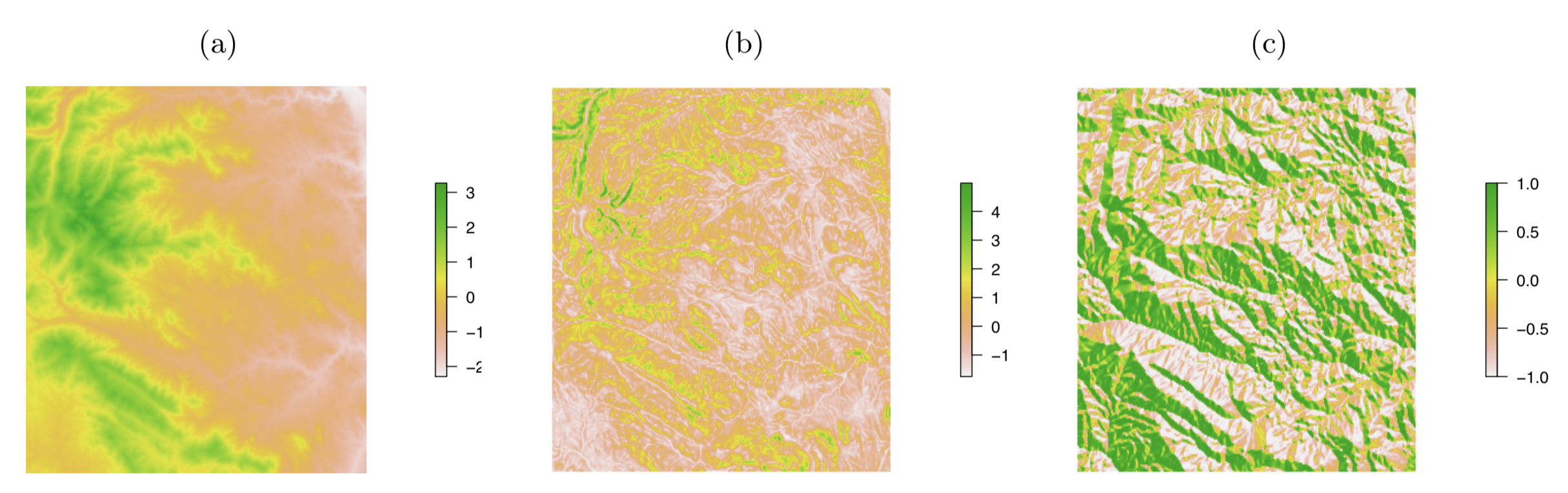}
  \caption{Maps of covariates in $\mathbf{W}$, including: a) elevation (standardized), b) slope (standardized), and c) solar exposure (ranges from $-1$ to $1$) in Colorado, USA.} 
  \label{fig:ml_covs}
\end{figure}

To demonstrate our empirical Bayes approach, we fit an EDE-based STPP model to a set of telemetry data from a mountain lion (\emph{Puma concolor}) in Colorado, USA.  These data comprise a set of 150 global positioning system (GPS) satellite fixes at a temporal resolution of 3 hours spanning a period of approximately 2.5 weeks (Figure~\ref{fig:ml_rs}).  
During this period our study individual moved several kilometers on a loop in the foothills northwest of Denver, CO as it exhibited normal life history behaviors for this species (Buderman et al., 2018; Hooten et al., 2019).  To improve our understanding of spatially heterogeneous motility and residence time, we specified a Bayesian model based on the conditional likelihood associated with EDE-based STPP model and multivariate normal prior for $\boldsymbol\theta$ with mean zero and diagonal covariance matrix with diagonal elements $0.1$, $1$, $1$, and $1$ (based on an intercept and three covariates) which induces a regularization on $\boldsymbol\theta$ and flattens the implicit prior on $\psi(\mathbf{s})$.  We used spatially explicit covariates that represent potentially important topographic resources for mountain lion movement, including standardized elevation and slope, as well as solar exposure (Figure~\ref{fig:ml_covs}). 
We estimated $\bar\delta(t_i)$ using a moving average of approximately 70 hours based on equation (\ref{eq:deltabar}) and obtained an availability sample of size $J-1=100$ for each position from a bivariate normal availability distribution implied by (\ref{eq:avlfcn}).  Using the procedure described in the previous section, we created an augmented binary data set and fit the EDE-based STPP model using a HMC algorithm (Appendix C) with 20,000 iterations and discarded the burn-in period of 1000 iterations.   

\begin{figure}[htp]
  \centering
  \includegraphics[width=5in, angle=0]{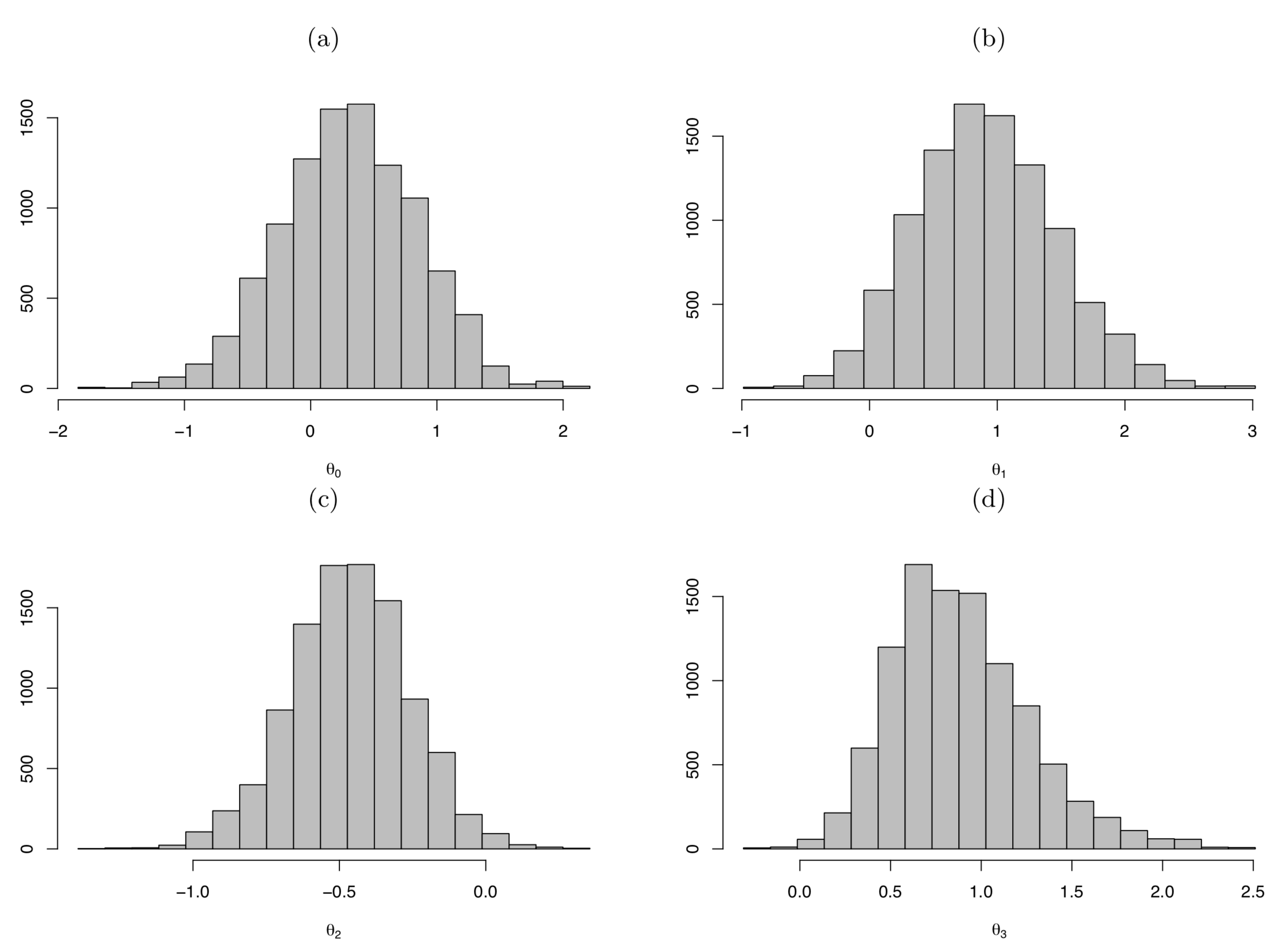}
  \caption{Marginal posterior distributions for motility coefficients $\boldsymbol\theta$ with: a) intercept, b) elevation coefficient, c) slope coefficient, and d) solar exposure coefficient.}
  \label{fig:ml_post}
\end{figure}

The results of our model fit to the mountain lion GPS data yielded marginal posterior distributions for the motility coefficients $\boldsymbol\theta$ shown in Figure~\ref{fig:ml_post}.    
We also computed the posterior mean of the derived quantity $r(\mathbf{s})$ in (\ref{eq:rs}) for the entire study area in units of hours per hectare (Figure~\ref{fig:ml_rs}).   

The results of our analysis suggest that the environmental covariates we used related to motility (and hence residence time) for the individual mountain lion data during the period of the study.  In particular, the GPS data suggested that the effect of both elevation and solar exposure had a positive relationship with motility whereas slope (i.e., steepness) had a negative relationship with motility (Figure~\ref{fig:ml_post}).  The posterior mean map of residence time confirms these findings and indicates that the individual remains longer in habitat with lower elevation as well as steeper and less exposed hillsides which generally consist of wetter, more densely forested areas.  By contrast, these results indicate that the mountain lion moves quickly through areas on mountaintops and ridges that are more exposed.       

From a management perspective, the movement of wildlife is often characterized spatially by movement corridors based on habitat preference or use (perhaps derived from conventional exponential resource selection analyses).  By considering the movement trajectories of wildlife in terms of a physically based, dynamic movement model, we can infer a variety of environmental conditions that may be important to conserve species and their natural movement patterns.  For example, in the case of the mountain lion we studied here, areas with greater residence time may be critical for one aspect of the life history of the animal, but areas with great motility may be important for transit between areas with higher residence time.  Our modeling framework allows managers to make inference on both aspects of wildlife movement behavior while using an analysis procedure that is intuitive and familiar. 

\section{Discussion}
Despite the rapidly growing popularity of discrete-time models (e.g., Morales et al., 2004) and slowly increasing popularity of stochastic differential equation (SDE) models for animal trajectories (e.g., Johnson et al., 2008b; Blackwell et al., 2016; Hooten and Johnson, 2017), point process models are still the predominant default method for obtaining animal movement inference given individual-based telemetry data.  Spatial and STPP models for animal telemetry data typically rely on exponential forms for the selection function $g(\mathbf{w}(\mathbf{s}(t_i)),\boldsymbol\theta)$ which only allow for inference on relative selection of resources.  

Garlick et al. (In Review) showed that a different form of selection function arises under the EDE that results from a first-principles perspective of animal movement.  We parameterized the selection component of the STPP in (\ref{eq:stpp}) based on a homogenized version of the EDE.  We linked the movement probability to the underlying environmental features that may influence movement using the logit-linear relationship $\text{logit}(\psi(\mathbf{s}))=\mathbf{w}'(\mathbf{s})\boldsymbol\theta$.  A natural characteristic of the EDE is that motility and residence time are inversely related, thus, we can easily make inference on residence time $r(\mathbf{s})$ as a derived quantity in the model.  This inference can be used in efforts to manage and conserve wildlife.

To fit the EDE-based STPP using computationally efficient algorithms, we derived the conditional regression likelihood that can be used to analyze data structures that are created using procedures that are standard practice in wildlife ecology.  While the conditional likelihood can be used in both maximum likelihood and Bayesian settings, we found it helpful to use Monte Carlo methods to fit a Bayesian version of the model because of the non-Gaussian joint posterior distributions that can result and to streamline the inference on derived quantities such as $r(\mathbf{s})$.   

The functional form of resource selection in point process models for animal movement has long been debated among ecologists.  While the exponential form of RSF is the most common by far, it is also limited in that it can only provide inference about relative selection.  Related to this, it is important to note that certain forms of RSFs can suffer from identifiability issues when estimating the parameters because the likelihood contains a ratio in which a globally multiplicative term cancels in the numerator and denominator of (\ref{eq:stpp}).  For example, the reason why an intercept is not included in the exponential RSF is because it will cancel.  Similarly, the parameters $\boldsymbol\theta$ in linear RSFs such as $g(\mathbf{w}(\mathbf{s}(t_i)),\boldsymbol\theta)=\mathbf{w}'(\mathbf{s}(t_i))\boldsymbol\theta$ and inverse linear RSFs such as $g(\mathbf{w}(\mathbf{s}(t_i)),\boldsymbol\theta)=1/\mathbf{w}'(\mathbf{s}(t_i))\boldsymbol\theta$ will result in the same likelihood if multiplied by a constant $c$.  As a result, this identifiability issue implies that the STPP cannot distinguish between $\boldsymbol\theta$ and $c\cdot \boldsymbol\theta$ for any $c\neq 1$ when using linear or inverse linear RSFs.  Consequently, if a more general RSF $g(\mathbf{w}(\mathbf{s}(t_i)),\boldsymbol\theta)$ is nearly linear or inverse linear in $\boldsymbol\theta$ on the study domain, the STPP model parameters may not all be fully identifiable.   

With the preceding note about identifiability in mind, overall, we showed that by combining the EDE-based STPP model with conditional regression approaches to fit the model, ecologists may be able to gain a new perspective on animal movement dynamics.  The computational approach we presented aligns with the most common way STPP models are fit to telemetry data in conventional step-selection analyses.  As part of ongoing research, we are assessing a suite of other PDEs for use in statistical models based on a similar procedure. 

\section*{Acknowledgements}
This research was funded by National Park Service Inventory and Monitoring Program and NSF DMS 1614392.  Any use of trade, firm, or product names is for descriptive purposes only and does not imply endorsement by the U.S. Government.

\section*{References}

\rf Anderson-Sprecher, R. and J. Ledolter.  (1991).  State-Space analysis of wildlife telemetry data. Journal of the American Statistical Association, 86: 596-602.

\rf Blackwell, P.G., M. Niu, M.S. Lambert, and S.D. LaPoint. (2016). Exact Bayesian inference for animal movement in continuous time.  Methods in Ecology and Evolution, 7: 184-195.

\rf Breslow, N.E. and W. Day. (1980). Statistical Methods in Cancer research: The analysis of case-control studies. Oxford University Press.  Oxford, UK.

\rf Brost, B.M., M.B. Hooten, E.M. Hanks, and R.J. Small. (2015). Animal movement constraints improve resource selection inference in the presence of telemetry error. Ecology, 96: 2590-2597. 

\rf Buderman, F.E., M.B. Hooten, M.W. Alldredge, E.M. Hanks, and J.S. Ivan. (2018). Time-varying predatory behavior is primary predictor of fine-scale movement of wildland-urban cougars. Movement Ecology, 6: 22. 

\rf Cagnacci, F., L. Boitani, R.A. Powell, and M.S. Boyce. (2010). Animal ecology meets {GPS}-based radiotelemetry:  A perfect storm of opportunities and challenges. Philosophical Transactions of the Royal Society B: Biological Sciences, 365: 2157-2162.

\rf Cooke, S.J., S.G. Hinch, M. Wikelski, R.D. Andrews, L.J. Kuchel, T.G. Wolcott, and P.J. Butler. (2004). Biotelemetry: A mechanistic approach to ecology. Trends in Ecology and Evolution, 19: 334-343.

\rf Cressie, N.A.C.  (1993).  Statistics for Spatial Data, Revised Edition.  Wiley.

\rf Cressie, N. and C. Wikle. 2011. Statistics for Spatio-Temporal Data. John Wiley and Sons, New York, New York, USA.

\rf Fortin, D., H.L. Beyer, M.S. Boyce, D.W. Smith, T. Duchesne, and J.S. Mao. (2005).  Wolves influence elk movements: Behavior shapes a trophic cascade in Yellowstone National Park. Ecology, 86: 1320-1330.

\rf Garlick, M.J., J.A. Powell, M.B. Hooten, and L. McFarlane. (2011). Homogenization of large-scale movement models in ecology. Bulletin of Mathematical Biology, 73: 2088-2108. 

\rf Garlick, M.J., J.A. Powell, M.B. Hooten, and L. McFarlane. (2014). Homogenization, sex, and differential motility predict spread of chronic wasting disease in mule deer in Southern Utah. Journal of Mathematical Biology, 69: 369-399. 

\rf Garlick, M.J., J.A. Powell, M.B. Hooten, and N. Brubaker. (In Review). Using homogenization to estimate random-walk motility from telemetry data in heterogeneous landscapes.  

\rf Gelfand, A.E. and S. Ghosh.  (2015).  Hierarchical Modeling.  In: Bayesian Theory and Applications.  Eds:  P. Damien, P. Dellaportas, N.G. Polson, and D.A. Stephens.  Oxford University Press.  

\rf Gelfand, A.E. and A.F. Smith. (1990). Sampling-based approaches to calculating marginal densities. Journal of the American Statistical Association, 85: 398-409.

\rf Gerber, B.D., M.B. Hooten, C.P. Peck, M.B. Rice, J.H. Gammonley, A.D. Apa, and A.J. Davis. (2019). Extreme site fidelity as an optimal strategy in an unpredictable and homogeneous environment. Functional Ecology, 33: 1695-1707.

\rf Haberman, R.  (2013).  Applied Partial Differential Equations with Fourier Series and Boundary Value Problems, 5th Edition.  Pearson.  

\rf Hanks, E.M., M.B. Hooten, and M. Alldredge. (2015). Continuous-time discrete-space models for animal movement. Annals of Applied Statistics, 9: 145-165. 

\rf Hefley, T.J., M.B. Hooten, R.E. Russell, D.P. Walsh, and J. Powell. (2017). When mechanism matters: Forecasting the spread of disease using ecological diffusion. Ecology Letters, 20: 640-650. 

\rf Hooten, M.B. and T.J. Hefley. (2019).  Bringing Bayesian Models to Life. Chapman \& Hall/CRC.

\rf Hooten, M.B. and D.S. Johnson. (2017). Basis function models for animal movement. Journal of the American Statistical Association, 112: 578-589.

\rf Hooten, M.B. and D.S. Johnson. (2019). Modeling Animal Movement. Gelfand, A.E., M. Fuentes, and J.A. Hoeting (eds). In: Handbook of Environmental and Ecological Statistics. Chapman \& Hall/CRC.

\rf Hooten, M.B., D.S. Johnson, B.T. McClintock, and J.M. Morales. (2017). Animal Movement: Statistical Models for Telemetry Data. Chapman \& Hall/CRC.

\rf Hooten, M.B., H.R. Scharf, and J.M. Morales. (2019). Running on empty: Recharge dynamics from animal movement data. Ecology Letters, 22: 377-389.  

\rf Hooten, M.B. and C.K. Wikle. (2008). A Hierarchical Bayesian non-linear spatio-temporal model for the spread of invasive species with application to the Eurasian Collared-Dove. Environmental and Ecological Statistics, 15: 59-70.

\rf Hooten, M.B. and C.K. Wikle. (2010). Statistical agent-based models for discrete spatio-temporal systems. Journal of the American Statistical Association, 105: 236-248.

\rf Hooten, M.B., F.E. Buderman, B.M. Brost, E.M. Hanks, and J.S. Ivan. (2016). Hierarchical animal movement models for population-level inference. Environmetrics, 27: 322-333.

\rf Johnson, D.S., D.L. Thomas, J.M. Ver Hoef, and A. Christ. (2008a). A general framework for the analysis of animal resource selection from telemetry data. Biometrics, 64: 968-976.

\rf Johnson, D.S., J.M. London, M.A. Lea, and J.W. Durban. (2008b). Continuous-time correlated random walk model for animal telemetry data. Ecology, 89: 1208-1215.

\rf Johnson, D.S., M.B. Hooten, and C.E. Kuhn. (2013). Estimating animal resource selection from telemetry data using point process models. Journal of Animal Ecology, 82: 1155-1164.

\rf Jonsen, I., J. Flemming, and R. Myers. (2005). Robust state-space modeling of animal movement data.  Ecology, 45: 589-598.

\rf Lele, S.R. (2009).  A new method for estimation of resource selection probability function.  The Journal of Wildlife Management, 73: 122-127.  

\rf Lele, S.R. and J.L. Keim.  (2006).  Weighted distributions and estimation of resource selection probability functions.  Ecology, 87: 3021-3028. 

\rf Lindgren, F., H. Rue, and J. Lindstrom. (2011).  An explicit link between Gaussian fields and Gaussian Markov random fields: The stochastic partial differential equation approach. Journal of the Royal Statistical Society: Series B, 73: 423-498.

\rf Logan, J.D.  (2015).  Applied Partial Differential Equations.  Springer-Verlag. 

\rf Lu, X., P.J. Williams, M.B. Hooten, J.A. Powell, J.N. Womble, and M.R. Bower. (In Press). Nonlinear reaction-diffusion process models improve inference for population dynamics. Environmetrics.

\rf Manly, B.F.J., L.L. McDonald, D.L. Thomas, T.L. McDonald, and W.P. Erickson. (2002). Resource Selection by Animals. Kluwer Academic Publishers, Dordrecht.

\rf McClintock, B.T. and T. Michelot. (2018). momentuHMM: R package for generalized hidden Markov models of animal movement.  Methods in Ecology and Evolution, 9: 1518-1530.

\rf Morales, J.M., D.T. Haydon, J. Frair, K.E. Holsinger, and J.M. Fryxell. (2004). Extracting more out of relocation data: Building movement models as mixtures of random walks. Ecology, 85: 2436-2445.

\rf Nathan, R., W.M. Getz, E. Revilla, M. Holyoak, R. Kadmon, D. Saltz, P.E. Smouse. (2008). A movement ecology paradigm for unifying organismal movement research. Proceedings of the National Academy of Science, 105: 19052-19059.

\rf Northrup, J.M., M.B. Hooten, C.R. Anderson, and G. Wittemyer. (2013). Practical guidance on characterizing availability in resource selection functions under a use-availability design. Ecology, 94: 1456-1464.

\rf Patil G. and C. Rao. (1977). The weighted distributions: a survey of their applications. In Applications of Statistics, P Krishnaiah (ed.). North Holland Publishing Company: Amsterdam, the Netherlands.

\rf Powell, J.A. and N.E. Zimmermann. (2004). Multiscale analysis of active seed dispersal contributes to resolving Reid's paradox. Ecology, 85: 490-506.

\rf Signer, J., J. Fieberg, and T. Avgar. (2019). Animal movement tools (amt): R package for managing tracking data and conducting habitat selection analyses.  Ecology and Evolution, 9: 880-890.

\rf Solymos, P. and S.R. Lele.  (2016).  Revisiting resource selection probability functions and single-visit methods: Clarification and extensions.  Methods in Ecology and Evolution, 7: 196-205.  

\rf Williams, P.J., M.B. Hooten, J.N. Womble, G.G. Esslinger, M.R. Bower, and T.J. Hefley. (2017). An integrated data model to estimate spatio-temporal occupancy, abundance, and colonization dynamics. Ecology, 98: 328-336.  

\rf Williams, P.J., M.B. Hooten, G.G. Esslinger, J.N. Womble, J. Bodkin, and M.R. Bower. (2019). The rise of an apex predator following deglaciation. Diversity and Distributions, 25: 895-908.

\rf Wikle, C.K. (2003). Hierarchical Bayesian methods for predicting the spread of ecological processes. Ecology, 84: 1382-1394.

\rf Zucchini, W. and I. MacDonald. (2009). Hidden Markov models for Time Series: An Introduction Using R. CRC Press. Boca Raton, Florida, USA.

\section*{Appendix A:  Multinomial Regression Approximation}
To show that a multinomial model results in the same likelihood as conditional logistic regression and thus can be used as an approximation to the STPP likelihood, we let $\mathbf{y}_i$ be the $J\times 1$ containing a first element equal to 1 and the remainder $J-1$ values equal to 0 for step $i$.  If we specify a multinomial model for the data such that $\mathbf{y}_i \sim \text{MN}(1,\boldsymbol\phi_i)$ with probabilities  
\begin{equation}  
  \phi_{ij} = \frac{g(\mathbf{w}_{ij},\boldsymbol\theta)}{\sum_{l=1}^J g(\mathbf{w}_{il},\boldsymbol\theta)} \;,
\end{equation} 
\noindent then the likelihood component for step $i$ is 
\begin{equation}
  [\mathbf{y}_i|\boldsymbol\theta] = \frac{1!}{1!\prod_{j=2}^J 0!} \phi_{i1}^1 \prod_{j=2}^J \phi_{ij}^0 \;,
\end{equation}
\noindent which yields the joint likelihood
\begin{equation}
  [\mathbf{Y}|\boldsymbol\theta] = \prod_{i=2}^n \frac{g(\mathbf{w}_{i1},\boldsymbol\theta)}{\sum_{j=1}^J g(\mathbf{w}_{ij},\boldsymbol\theta)}.  
  \label{eq:mn_lik}
\end{equation}
\noindent Note that the likelihood resulting from the multinomial model for $\mathbf{y}_i$ in (\ref{eq:mn_lik}) coincides with the conditional logistic regression likelihood in (\ref{eq:condlik}).

\section*{Appendix B:  EDE Homogenization}
To derive the EDE-based STPP formulation that results in the selection and availability functions in (\ref{eq:selfcn}) and (\ref{eq:avlfcn}), we begin from the diffusion-based Fokker-Plank equation (i.e., the EDE) in (\ref{eq:EDE}) and then derive the associated homogenized EDE using the method of multiple scales.  Garlick et al. (2011) showed that the plain diffusion equation 
\begin{equation}
  \frac{\partial u(\mathbf{s},t)}{\partial t}=\bar\delta(\mathbf{s})\left(\frac{\partial^2}{\partial s^2_1}+\frac{\partial^2}{\partial s^2_2}\right)u(\mathbf{s},t) \;,
  \label{eq:hede}
\end{equation}
can be solved numerically on a coarser spatio-temporal scale which facilitates much faster algorithms and resulting statistical inference (Hooten et al., 2013).  The homogenized motility function $\bar\delta(\mathbf{s})$ in (\ref{eq:hede}) is the harmonic mean of $\delta(\mathbf{s})$ over the coarser scale.  Thus, to return to the EDE on the fine scale, we divide the homogenized process by the fine scale motility function $p(\mathbf{s},t)=u(\mathbf{s},t)/\delta(\mathbf{s})$.  This upscaling strategy substantially reduces computing requirements to implement a statistical model containing the EDE (Hooten et al., 2013).       

A secondary benefit, and one that we exploit here, is that the homogenized EDE in (\ref{eq:hede}) also allows us to solve for $u(\mathbf{s},t)$ given a point source at the previous time $t-\Delta t$.  Following Haberman (2004) and Logan (2015), for previous position $\mathbf{s}(t-\Delta t)$, the fundamental solution is
\begin{equation}   
  u(\mathbf{s},t)=|2\pi 2\bar\delta(\mathbf{s}) \Delta t\mathbf{I}|^{-\frac{1}{2}}e^{-\frac{1}{2}(\mathbf{s}-\mathbf{s}(t-\Delta t))'(2\bar\delta(\mathbf{s}) \Delta t\mathbf{I})^{-1}(\mathbf{s}-\mathbf{s}(t-\Delta t))} \;.
\end{equation}   
Then, because $p(\mathbf{s},t)$ can be recovered by dividing $u(\mathbf{s},t)$ by the motility function, we have 
\begin{equation}   
  p(\mathbf{s},t)\propto \frac{1}{\delta(\mathbf{s})\Delta t}e^{-\frac{1}{2}(\mathbf{s}-\mathbf{s}(t-\Delta t))'(2\bar\delta(\mathbf{s}) \Delta t\mathbf{I})^{-1}(\mathbf{s}-\mathbf{s}(t-\Delta t))} \;,
\end{equation}   
\noindent which matches the conditional distribution for $\mathbf{s}(t)$ in (\ref{eq:pss}).  Thus, the homogenized fundamental solution to the EDE can serve as the point process model in a step selection analysis.     

\section*{Appendix C:  Hamiltonian Monte Carlo Algorithm}
In our implementation of the HMC algorithm, we define the Hamiltonian function as a function of position $\bm{\theta}$ (in parameter space) and velocity $\bm{v}$ in the following,
\begin{equation}
  h(\bm{\theta}, \mathbf{v}) = -\log[\bm{\theta}|\cdot] - \log[\mathbf{v}] \;, \label{eq:h}
\end{equation}
where $[\bm{\theta}|\cdot]$ is the full-conditional distribution of $\bm{\theta}$ up to a multiplicative constant and $[\mathbf{v}] = \text{N}\left(\mathbf{0}, \bm{\Sigma}_v\right)$. Following the conditional likelihood derivation, the full conditional distribution for $\bm{\theta}$ is
\begin{align*}
    [\bm{\theta}|\cdot] &\propto \left[\mathbf{y}\left|\bm{\theta}, \left\{\sum_{j = 1}^{J_i}y_{ij} = 1, \forall i\right\}\right.\right][\bm{\theta}] ,\\
    &= \prod_{i = 2}^n\frac{g_{i1}}{\sum_{j = 1}^{J_i}g_{ij}} \cdot \text{N}\left(\bm{\mu}_\theta, \bm{\Sigma}_\theta\right),
\end{align*}
where $g_{ij} \equiv g\left(\mathbf{w}_{ij}, \bm{\theta}\right)$ and $\bm{\mu}_\theta$ and $\bm{\Sigma}_\theta$ are hyperparameters for $\boldsymbol\theta$. The Hamiltonian trajectories are controlled by partial derivatives of the Hamiltonian function in (\ref{eq:h}),
\begin{align}
    \frac{d\mathbf{v}(\tau)}{d\tau} &= -\frac{\partial h(\bm{\theta}, \mathbf{v})}{\partial \bm{\theta}} = \sum_{i = 2}^n\left(\frac{\bigtriangledown g_{i1}}{g_{i1}} - \frac{\sum_{j = 1}^{J_i}\bigtriangledown g_{ij}}{\sum_{j = 1}^{J_i}g_{ij}}\right) - \bm{\Sigma}_{\theta}^{-1}\left(\bm{\theta} - \bm{\mu}_\theta\right), \label{eq:h1} \\
    \frac{d\bm{\theta}(\tau)}{d\tau} &= \frac{\partial h(\bm{\theta}, \mathbf{v})}{\partial \mathbf{v}} = \bm{\Sigma}_v^{-1}\left(\mathbf{v} - \mathbf{0}\right), \label{eq:h2}
\end{align}
where $\bigtriangledown g_{ij} \equiv \partial g\left(\mathbf{w}_{ij}, \bm{\theta}\right)/\partial \bm{\theta}$ and $\boldsymbol\Sigma_v\equiv 3\cdot\mathbf{I}$ is set as a tuning parameter. When we use the link function $g\left(\mathbf{w}_{ij}, \bm{\theta}\right) = \left(\text{logit}^{-1}\left(\mathbf{w}'_{ij}\bm{\theta}\right)\right)^{-1} = \frac{1 + \exp\left(\mathbf{w}'_{ij}\bm{\theta}\right)}{\exp\left(\mathbf{w}'_{ij}\bm{\theta}\right)}$, the gradient function $\partial h(\bm{\theta}, \mathbf{v})/\partial \bm{\theta}$ is evaluated as
$$
\frac{\partial h(\bm{\theta}, \mathbf{v})}{\partial \bm{\theta}} = -\sum_{i = 2}^n\left(\frac{-\mathbf{w}_{i1}}{1 + \exp\left(\mathbf{w}'_{i1}\bm{\theta}\right)} - \frac{\sum_{j = 1}^{J_i}-\exp\left(-\mathbf{w}'_{ij}\bm{\theta}\right)\mathbf{w}_{ij}}{\sum_{j = 1}^{J_i}\frac{1 + \exp\left(\mathbf{w}'_{ij}\bm{\theta}\right)}{\exp\left(\mathbf{w}'_{ij}\bm{\theta}\right)}}\right) + \bm{\Sigma}_\theta^{-1}\left(\bm{\theta} - \bm{\mu}_\theta\right).
$$

Based on the Hamiltonian system in (\ref{eq:h1}) and (\ref{eq:h2}), the associated leap frog algorithm in computing time $\tau$ with discretization $\Delta \tau$ is
\begin{enumerate}
    \item Choose initial velocity $\mathbf{v}(0)$;
    \item Update the velocity a half step in time using
    $$
    \mathbf{v}\left(\tau + \frac{\Delta \tau}{2}\right) = \mathbf{v}(\tau) - \frac{\Delta \tau}{2}\frac{\partial h(\bm{\theta}(\tau), \mathbf{v}(\tau))}{\partial \bm{\theta}};
    $$
    \item Update the position using 
    $$
    \bm{\theta}(\tau + \Delta \tau) = \bm{\theta}(\tau) + \Delta \tau\frac{\partial h\left(\bm{\theta}(\tau), \mathbf{v}\left(\tau + \frac{\Delta \tau}{2}\right)\right)}{\partial \mathbf{v}};
    $$
    \item Update the velocity again using
    $$
    \mathbf{v}(\tau + \Delta \tau) = \mathbf{v}\left(\tau + \frac{\Delta \tau}{2}\right) - \frac{\Delta \tau}{2}\frac{\partial h\left(\bm{\theta}(\tau + \Delta \tau), \mathbf{v}\left(\tau + \frac{\Delta \tau}{2}\right)\right)}{\partial \bm{\theta}};
    $$
    \item Let $\tau = \tau + \Delta \tau$, go to 2 and repeat until the end of computing time period for each update.
\end{enumerate}
In practice, we tuned the HMC algorithm such that $\Delta \tau=0.05$ and the maximum $\tau$ was 10 to yield a well-mixed Markov chain for $\boldsymbol\theta$.  

\end{document}